\documentclass{article}

\usepackage{spconf,amsmath,graphicx,hyperref,multirow}

\title{Windformer: Bi-directional Long-Distance Spatio-Temporal Network for Wind Speed Prediction}
\name{Xuewei Li$^{1, 2, 3, 4}$, Zewen Shang$^{1, 2, 3}$, Zhiqiang Liu$^{1, 2, 3}$, Jian Yu$^{1, 2, 3}$, Wei Xiong$^{5, *}$\thanks{*Corresponding author.}, Mei Yu$^{1, 2, 3, 4}$ \thanks{This work is supported by National Natural Science Foundation of China (Grant No. 61976155).}}

\address{
	$^{1\;}$College of Intelligence and Computing, Tianjin University, Tianjin, 300350, China.
	\\$^{2\;}$Tianjin Key Laboratory of Cognitive Computing and Application, Tianjin, 300350, China.
	\\$^{3\;}$Tianjin Key Laboratory of Advanced Networking, Tianjin, 300350, China.
	\\$^{4\;}$Tianjin International Engineering Institute, Tianjin University, Tianjin, 300350, China.
    \\$^{5\;}$TCU School of Civil Engineering, Tianjin Chengjian University, Tianjin 300384, China
}

\begin{document}
%

\maketitle

\begin{abstract}
Wind speed prediction is critical to the management of wind power generation. Due to the large range of wind speed fluctuations and wake effect, there may also be strong correlations between long-distance wind turbines. This difficult-to-extract feature has become a bottleneck for improving accuracy. History and future time information includes the trend of airflow changes, whether this dynamic information can be utilized will also affect the prediction effect. In response to the above problems, this paper proposes Windformer. First, Windformer divides the wind turbine cluster into multiple non-overlapping windows and calculates correlations inside the windows, then shifts the windows partially to provide connectivity between windows, and finally fuses multi-channel features based on detailed and global information. To dynamically model the change process of wind speed, this paper extracts time series in both history and future directions simultaneously. Compared with other current-advanced methods, the Mean Square Error (MSE) of Windformer is reduced by 0.5\% to 15\% on two datasets from NERL.

\end{abstract}
\section{Introduction}
\label{sec:intro}

In the context of the global shift towards a low-carbon energy structure, wind energy has emerged as one of the most extensively developed and applied renewable energy sources. Wind speed prediction can improve the control foresight and safety of wind turbines, bringing significant safety value and economic benefits. Deep learning has achieved considerable results in the field of wind energy prediction, such as AGRU \cite{niu2020wind}, TACBiLSTMD \cite{ma2022hybrid}, EMD-LSTM-ARIMA \cite{liu2021application} and other RNN methods, FC-CNN \cite{yu2019scene}, U-ConvRes \cite{bastos2021u} and other CNN-like methods, as well as CNN-GRU \cite{kou2020deep}, STNN \cite{liu2020probabilistic}, etc. that combine the two.

\begin{figure}[htb]

\begin{minipage}[b]{0.35\linewidth}
  \centering
  \centerline{\includegraphics[width=0.9\linewidth]{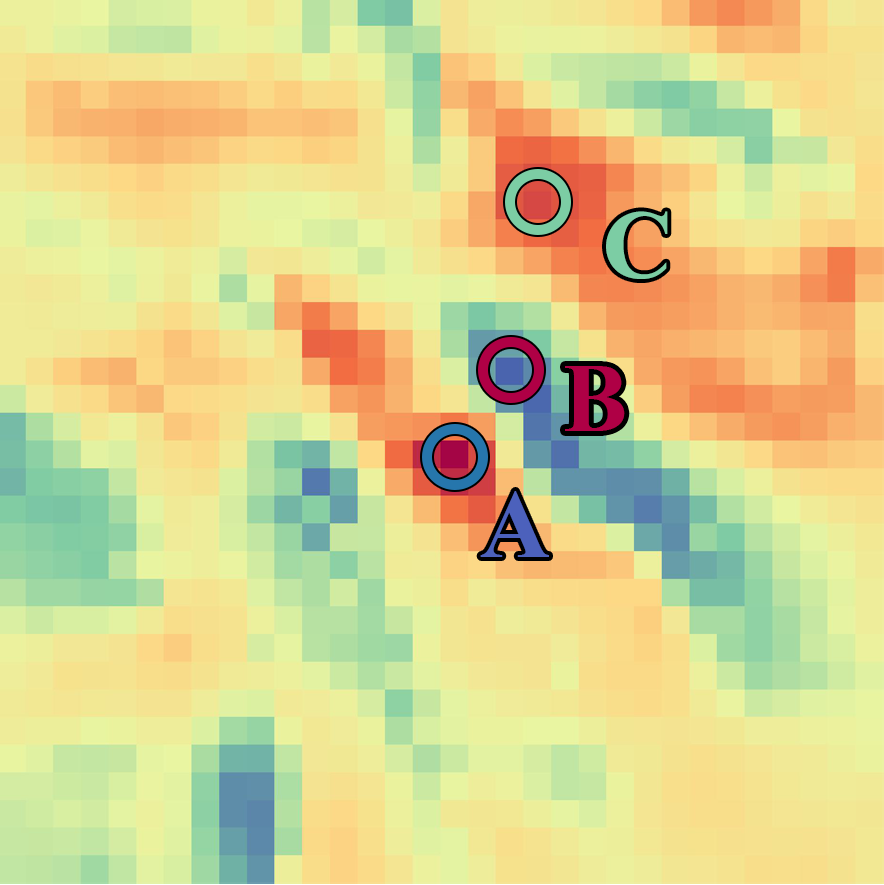}}
  \centerline{(a) MIC}  
\end{minipage}
\begin{minipage}[b]{0.65\linewidth}
  \centering
  \centerline{\includegraphics[width=0.9\linewidth]{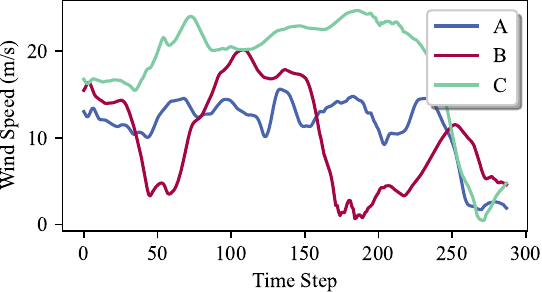}}
  \centerline{(b) Wind speed curve of three turbines} 
\end{minipage}
\caption{Example of long-distance correlations. (a) shows the Maximal Information Coefficients between the wind speeds of wind turbine A located at the center and the wind speeds of each other wind turbine on a randomly selected day. }
\label{corr}
\vspace{-0.4cm}
\end{figure}

It can be observed from Figure \ref{corr} that despite wind turbine C being at a long-distance from wind turbine A, it still exhibits a notable correlation compared to the nearby wind turbine B. This suggests that the wind speeds of the turbines are not only correlated with those in their immediate vicinity but also with wind speeds at certain long-distance intervals, emphasizing the significance of these relationships.

The self-attention mechanism was first proposed in Transformer \cite{vaswani2017attention}. Later, Vit \cite{dosovitskiy2020vit} integrated Transformer into CV tasks and achieved good results, especially in the extraction of global long-distance dependencies. Subsequently, Vit variants such as PVT \cite{wang2021pvtv2}, Conformer \cite{peng2021conformer}, Swin Transformer \cite{liu2021swin} appeared continuously. On this basis, this paper proposes a model suitable for dealing with wind speed problems, called Windformer. Windformer divides the wind turbines into multiple non-overlapping windows, performs correlation calculations for the wind turbines in the windows, and then shifts the windows to provide connectivity between windows. This article also designs the Channel Fusion Module to assign different weights to the data of different channels in the model to emphasize more important features. Then, Windformer uses Bi-ConvGRU to improve the ability to extract bidirectional timing information.

In summary, our main contributions are listed as follows.

(1) This paper proposes a Spatial Feature Extraction Module based on Sub-cluster Corr Blocks, which divides the wind turbine cluster into multiple sub-clusters and enables the dynamic extraction of spatial correlations over long-distance. This paper also introduces the Global-detail Fusion Module, which refines features by incorporating detailed and global information.

(2) A Temporal Feature Extraction Module based on forward and reverse bidirectional timing paths is proposed, which captures the temporal correlation of time steps based on past and future dynamic information.

(3) Windformer is experimentally validated on two large-scale wind turbine clusters, each containing over eight hundred wind turbines. The results show that Windformer performs better than other current-advanced methods in different tasks and reduces the MSE by 0.5\% to 10\%. Our code will be released at \\ \href{https://github.com/szwszwszw123/Windformer}{https://github.com/szwszwszw123/Windformer}

\section{Proposed Method}
\label{sec:format}

\begin{figure*}[t]

\begin{minipage}[b]{1\linewidth}
  \centering
  \centerline{\includegraphics[width=18cm]{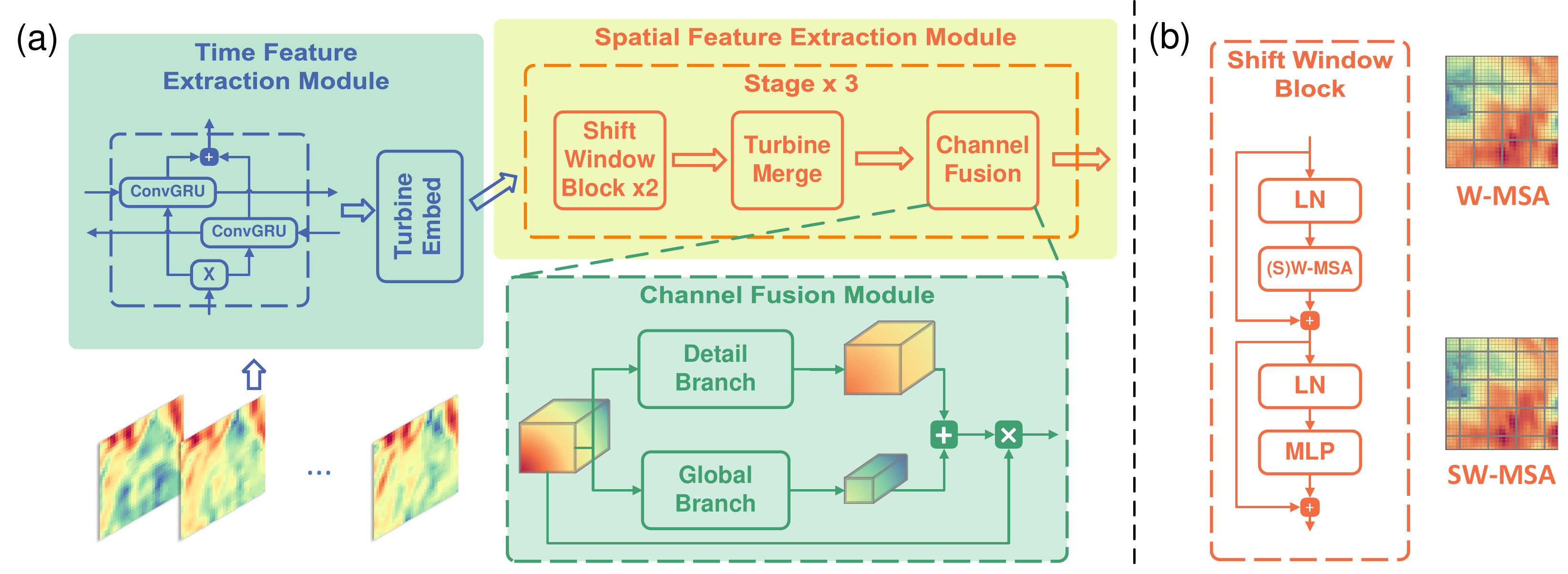}}
\end{minipage}
%

%
\caption{(a):Architecture of Windformer. (b): Internal Architecture of Shift Window Block. A stage contains two consecutive Shift Window Blocks, with the first using W-MSA, and the second using SW-MSA to provide connectivity for the window.}
\label{architecture}
\end{figure*}

\subsection{Problem Definition}

Windformer utilizes historical feature data from wind turbine cluster to predict 
future wind speeds. Following the concept introduced by Scene \cite{yu2019scene}, this paper embeds meteorological data from multiple wind turbines at a certain moment into a two-dimensional grid, referred to as a Scene denoted as $S^{t} \in R^{F\times H \times W}$. Here, $t$ represents the time step, $F$ signifies the number of input features, and $H, W$ represent the height and width of the grid, respectively. The model's input consists of multiple time steps, namely
$S = \{ S^{1}, S^{2}, ..., S^{T} \} \in R^{T \times F \times H \times W}$, where $T$ denotes the number of time steps. The output yields predicted results $\hat{Y} \in R^{L}$ for each wind turbine, where $L$ represents the number of wind turbines. The complete prediction process can be defined as follows:
\begin{equation}\label{...}
\hat{Y} = Windformer (S)
\end{equation}

The model architecture of Windformer is illustrated in Figure 
 \ref{architecture}(a). 

\subsection{Time Feature Extraction Module}
\label{sec:TFEM}
In wind speed prediction, the features from each time step are indispensable and must be considered collectively. In Recurrent Neural Network, each time step can only access historical features and not future ones, limiting the model's expressive capability. Therefore, Windformer employs the Bi-ConvGRU to handle input scenes, maximizing the utilization of bidirectional information. This method preserves spatial features while extracting temporal features.

GRU is a variant of LSTM. ConvGRU\cite{gru} modifies the fully connected layers in GRU to convolutional layers. To effectively utilize bidirectional information, this paper integrates the forward and backward ConvGRUs, forming a Bi-ConvGRU. This integration concatenates the hidden states of both directions at each time step and outputs them together, as shown in the equation below: 

\begin{equation}
\begin{aligned}
\overrightarrow{H_t}&=\overrightarrow{GRU}(X_t,H_{t-1}),t\in[1,T], \\
\overleftarrow{H_t}&=\overleftarrow{GRU}(X_t,H_{t+1}),t\in[1,T], \\
H_t&=[\overrightarrow{H_t},\overleftarrow{H_t}] \\
\end{aligned}
\end{equation}

where $\overrightarrow{H_t}$,$\overleftarrow{H_t}$ represent the hidden states from history and future, respectively. Compared with ConvGRU, Bi-ConvGRU can extract richer time series features from history and future time steps, and the extracted features are convenient for subsequent Spatial Feature Extraction Modules to process.

Following the Bi-ConvGRU, Windformer employs the Turbine Embed Module to independently extract features for each wind turbine. This involves utilizing a linear layer to transform the features of a wind turbine at different time steps into a vector. 
Subsequently, this vector is input into the Spatial Feature Extraction Module.

\subsection{Spatial Feature Extraction Module}
\subsubsection{Shift Window Block}

Drawing inspiration from Liu et al. \cite{liu2021swin}, this paper adopts the concept of shifted windows to extract spatial features. The Spatial Feature Extraction Module comprises three stages, each composed of two consecutive Shift Window Blocks, a Turbine Merge Module, and a Channel Fusion Module.

Self-attention has less inductive bias and can handle wind turbines at different distances equally, thus being able to handle both short-distance and long-distance correlations. Based on this background, this paper uses Window Multi-Head self-attention (W-MSA) and Shift Window Multi-Head self-attention (SW-MSA) \cite{liu2021swin} in Shift Window Block to capture spatial correlation. W-MSA divides the wind turbine into multiple non-overlapping windows and performs self-attention calculations inside the windows.

Utilizing only W-MSA would lead to a lack of continuity in information between different windows. Each wind turbine could only observe other wind turbines within its own window, making it difficult to extract wind speed information from long-distance. SW-MSA addresses this issue effectively by introducing a shift in the windows, establishing connectivity between adjacent windows, as depicted in Figure \ref{architecture} (b).

To further enhance the model's ability to represent wind speed features and prevent gradient explosion, the Shift Window Block also incorporates a Multi-Layer Perceptron followed by LayerNorm. Subsequently, two residual connections are introduced, as depicted in Figure \ref{architecture} (b).

\subsubsection{Turbine Merge Module}

After passing through two Shift Window Blocks, the generated features are input into the Turbine Merge Module. The Turbine Merge Module aggregates $2 \times 2$ adjacent data points into one, allowing the model to extract features of different granularities.

\subsubsection{Channel Fusion Module}

The Shift Window Block primarily focuses on modeling internal spatial relationships within the scene, but its modeling of multi-channel information across scenes is insufficient. As the stages deepen, the amount of information contained in an individual data point increases, leading to a growing number of channels. Merely linearly combining these channel features will cause the loss of feature information, so more detailed fusion modeling of different regions of different channels is required.

To address the aforementioned concern, Windformer employs the Channel Fusion Module at the end of each stage to further refine channel features and enhance their expressive capability. Drawing inspiration from the method by Dai et al. \cite{dai2021attentional}, this paper incorporates detailed context into the SENet \cite{hu2018squeeze} to combine detailed and global wind speed information. The Channel Fusion Module comprises two branches: Global and Detail. In Global Branch, the feature map is firstly pooled globally, and then the global wind speed information is obtained through two linear layers. In Detail Branch, the feature map is convolved twice, and finally, the global and detailed wind speed information is added.
Specifically, the detailed information $D(X)$ and the global information $G(X)$ can be computed as follows:

\begin{equation}
\begin{aligned}
X'   &= Pool(X), \\
D(X) &= BN(Conv(ReLU(BN(Conv(X))))), \\
G(X) &= BN(Linear(ReLU(BN(Linear(X'))))
\end{aligned}
\end{equation}
where BN refers to the BatchNorm layer. After computing $G(X)$ and $D(X)$, the enhanced features can be expressed as follows:

\vspace{-0.4cm}
\begin{equation}\label{...}
\begin{aligned}
X' = X \otimes M(X) =  X \otimes \sigma (L(X) + G(X))
\end{aligned}
\end{equation}

where $\otimes$ denotes element-wise multiplication and $\sigma$ represents the Sigmoid function.

\section{Experiment and Analysis}
\label{sec:pagestyle}

\subsection{Problem Definition}

\subsubsection{Datasets and Model}


This paper utilizes two dense wind turbine clusters from NREL \footnote{https://www.nrel.gov} as our datasets. The two clusters include 850 and 878 wind turbines respectively, with a grid resolution of 2km $\times$ 2km. This paper set prediction tasks for time intervals of 30 minutes, 60 minutes, and 90 minutes to assess the model's generalization capability. The input features include five meteorological features: wind speed, wind direction, air pressure, temperature, and air density. 


During model training, this paper set the batch size to 256 and utilizes the AdamW optimizer with a learning rate of $4\times10^{-3}$ and a weight decay of $10^{-4}$.

\subsubsection{Experimental Results and Comparison}


\begin{table}[tb]
\scriptsize
  \begin{center}
    \caption{Quantitative Evaluation Results on Dataset1. To adapt to the single-step prediction task and Scene size, this paper makes partial adjustments to other deep learning models, removing Reverse scheduled sampling in PredRNN and setting the Inception kernel size in SimVP to 3 and 5.}
    \label{dataset1_table}
    \resizebox*{\linewidth}{!}{
    \begin{tabular}{c|c|c|c|c|c|c}
    \hline
      \multirow{2}{*}{Model} & \multicolumn{3}{c|}{MSE} & \multicolumn{3}{c}{MAE}   \\ \cline{2-7}
       & 30 min & 60 min & 90 min & 30 min & 60 min & 90 min \\  \hline
              XGBR        & 2.597 & 4.977 & 6.945 & 0.999 & 1.521 & 1.880 \\ 
       KNN \cite{becker2017completion}  & 2.881 & 5.638 & 7.960 & 1.069 & 1.645 & 2.039 \\  
       SVR \cite{chen2014short} \          & 3.114 & 5.669 & 7.688 & 1.029 & 1.566 & 1.926 \\ 
     RegressionTree \cite{torres2019regression} & 2.989 & 5.440 & 7.400 & 1.118 & 1.619 & 1.961 \\ 
    STNN \cite{liu2020probabilistic} (2020)     & 1.287 & 3.249 & 4.966 & 0.702 & 1.212 & 1.566 \\ 
    MAU \cite{chang2021mau} (2021)  & 1.365 & 3.131 & 4.711 & 0.738 & 1.712 & 1.536\\ 
    PredRNN v2 \cite{wang2022predrnn} (2022)  & 1.282 & 3.150 & 4.703 & 0.700 & 1.200 & 1.545 \\ 
    SimVP \cite{gao2022simvp} (2022)    & 1.233 & 3.005 & 4.586 & 0.698 & 1.168 & 1.505 \\ \hline
    Windformer     & \textbf{1.197} & \textbf{2.914} & \textbf{4.536} & \textbf{0.681} & \textbf{1.149} & \textbf{1.491} \\ \hline
    \end{tabular}
}
  \end{center}
  \vspace{-0.8cm}
\end{table}

\begin{table}[tb]
\scriptsize
  \begin{center}
    \caption{Quantitative Evaluation Results on Dataset2}
    \label{dataset2_table}
    \resizebox*{\linewidth}{!}{
    \begin{tabular}{c|c|c|c|c|c|c}
    \hline
      \multirow{2}{*}{Model} & \multicolumn{3}{c|}{MSE}  & \multicolumn{3}{c}{MAE}   \\ \cline{2-7}
       & 30 min & 60 min & 90 min & 30 min & 60 min & 90 min \\  \hline
    XGBR         & 0.906 & 1.604 & 2.251 & 0.584 & 0.845 & 1.050 \\ 
       KNN \cite{becker2017completion} & 0.988 & 1.781 & 2.532 & 0.619 & 0.904 & 1.130 \\  
       SVR \cite{chen2014short}        & 0.984 & 1.753 & 2.448 & 0.59  & 0.858 & 1.066  \\ 
     RegressionTree \cite{torres2019regression} & 0.975 & 1.710 & 2.387 & 0.624 & 0.886 & 1.093 \\ 
STNN \cite{liu2020probabilistic}(2020)   & 0.472 & 0.962 & 1.550 & 0.416 & 0.633 & 0.838 \\ 
MAU \cite{chang2021mau} (2021)  & 0.467 & 0.917 & 1.389 & 0.415 & 0.608 & 0.799\\ 
PredRNN v2 \cite{wang2022predrnn}(2022) & 0.484 & 0.957 & 1.517 & 0.423 & 0.637 & 0.853 \\ 
SimVP \cite{gao2022simvp}(2022)  & 0.435 & 0.892 & 1.324 & 0.396 & 0.605 & 0.764 \\ \hline
       Windformer   & \textbf{0.430} & \textbf{0.883} & \textbf{1.318} & \textbf{0.396} & \textbf{0.592} & 
\textbf{0.761}
 \\ \hline
    \end{tabular}
    }
  \end{center}
  \vspace{-0.6cm}
\end{table}

Table \ref{dataset1_table} \& \ref{dataset2_table} provides a quantitative comparison of errors for various methods. Taking the 60-minute prediction task as an example, for two datasets, Windformer achieves MSE values of 2.914 and 0.883, respectively. Compared to machine learning methods, the MSE reduction for the two datasets ranges from 41.449\% to 48.596\%, and 44.932\% to 50.411\%, respectively. Moreover, as the prediction horizon extends, the reduction in errors becomes more pronounced, emphasizing the significance of extracting spatial correlations.

When compared to other current-advanced methods, the MSE of Windformer is reduced by 0.5\% to 15\%. This demonstrates that the Shift Window Block efficiently extracts correlations among wind turbines at varying distances. 

\begin{figure}[tb]

\begin{minipage}[b]{1.0\linewidth}
  \centering
  \centerline{\includegraphics[width=8.5cm]{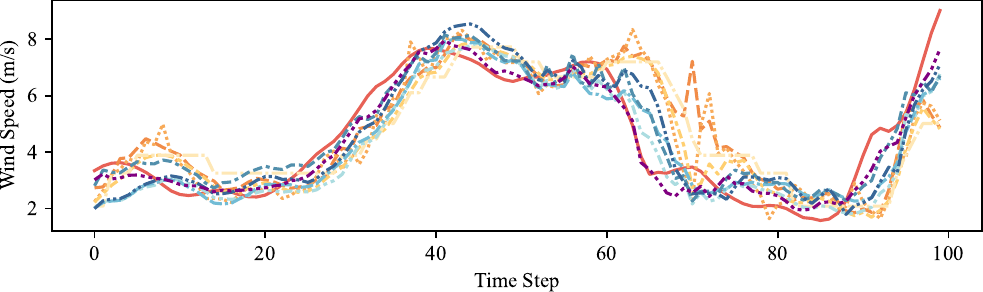}}
\end{minipage}

\begin{minipage}[b]{1.0\linewidth}
  \centering
  \centerline{\includegraphics[width=8.5cm]{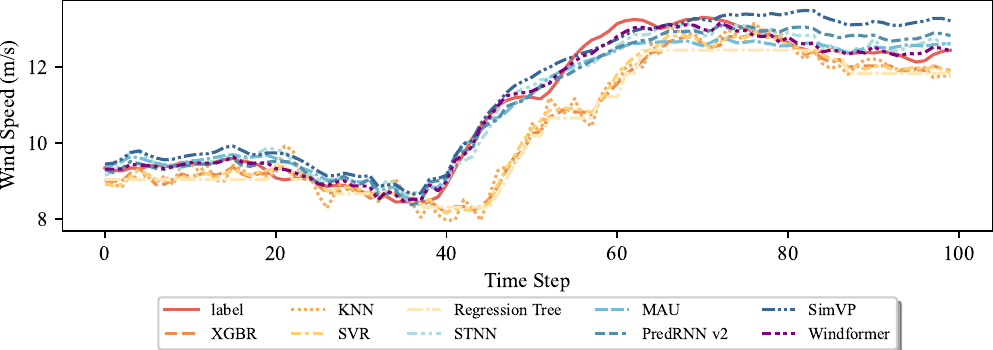}}
\end{minipage}
\vspace{-0.4cm}
\caption{Prediction curves of different methods on a randomly selected wind turbine on two datasets.}
\label{curve}

\end{figure}

It can be observed from Figure \ref{curve} that the prediction curve of Windformer is highly consistent with the real result, and the stability is stronger. The prediction curve of other methods has obvious hysteresis, while the curve of Windformer is highly consistent with the label. 

\subsubsection{Ablation Experiment}

In this study, this paper conducted ablation experiments on various modules using Dataset 1. First, this paper investigates the effectiveness of the Time Feature Extraction Module. It can be observed from Table \ref{TFEM} that Bi-ConvGRU achieved the best performance, and the improvement is obvious compared to ConvGRU and Bi-GRU. This illustrates the need to consider spatial features in the temporal feature extraction stage, and also illustrates the importance of bidirectional temporal feature extraction.

\begin{table}[h]
\scriptsize
  \begin{center}
  \vspace{-0.4cm}
    \caption{Ablation study of Time Feature Extraction Module. }
    \label{TFEM}
    \resizebox*{\linewidth}{!}{
    \begin{tabular}{c|c|c|c|c|c|c}
    \hline
      \multirow{2}{*}{Module} & \multicolumn{3}{c|}{MSE} & \multicolumn{3}{c}{MAE}   \\ \cline{2-7}
       & 30 min & 60 min & 90 min & 30 min & 60 min & 90 min \\  \hline
       Empty Module        & 1.273 & 3.124 & 4.754 & 0.714 & 1.196 & 1.726 \\  
       Bi-ConvRNN   & 1.214 & 3.013 & 4.699 & 0.686 & 1.162 & 1.518 \\  
       Bi-ConvLSTM  & 1.217 & 2.920 & 4.552 & 0.69 & 1.15 & 1.495 \\ 
       Bi-GRU      & 1.246 & 2.982 & 4.761 & 0.683 & 1.158 & 1.629  \\
       ConvGRU      & 1.230 & 3.019 & 4.566 & 0.692 & 1.173 & 1.515  \\
       Bi-ConvGRU   & \textbf{1.197} & \textbf{2.914} & \textbf{4.536} & \textbf{0.681} & \textbf{1.149} & \textbf{1.491} \\ \hline
    \end{tabular}
    }
    \vspace{-0.8cm}
  \end{center}
\end{table}

Secondly, this paper studies the effectiveness of the Spatial Feature Extraction Module. It can be observed from Table \ref{SFEM} that the error of the Shift Window Block is significantly lower than that of the CNN Module and the Empty Module. Removing the shift window (using two consecutive W-MSA) also reduces the prediction accuracy, which is a testament to the model's ability to capture long-distance correlations.

\begin{table}[h]
\scriptsize
  \begin{center}
  \vspace{-0.4cm}
    \caption{Ablation study of Spatial Feature Extraction Module.}
    \label{SFEM}
    \resizebox*{\linewidth}{!}{
    \begin{tabular}{c|c|c|c|c|c|c}
    \hline
      \multirow{2}{*}{Module} & \multicolumn{3}{c|}{MSE} & \multicolumn{3}{c}{MAE}   \\ \cline{2-7}
       & 30 min & 60 min & 90 min & 30 min & 60 min & 90 min \\  \hline
       Empty Module         & 1.619 & 3.760 & 5.478 & 0.795 & 1.314 & 1.663 \\  
       CNN Module          & 1.235 & 2.985 & 4.626 & 0.701 & 1.171 & 1.513 \\  
       Window          & 1.352 & 3.264 & 4.729 & 0.752 & 1.267 & 1.572 \\  
       Shift Window  & \textbf{1.197} & \textbf{2.914} & \textbf{4.536} & \textbf{0.681} & \textbf{1.149} & \textbf{1.491} \\  \hline
    \end{tabular}
    }
    \vspace{-0.8cm}
  \end{center}
\end{table}

Subsequently, this paper ablates each Branch in the Channel Fusion Module. From the experiment, it can be observed that the removal of Global Branch or Detail Branch will affect the accuracy of the prediction, which shows that both global feature and detail feature will affect the prediction results, and need to be paid attention to at the same time.

\begin{table}[h]
\scriptsize
  \begin{center}
  \vspace{-0.4cm}
    \caption{Ablation study of Channel Fusion Module. }
    \label{TFEM}
    \resizebox*{\linewidth}{!}{
    \begin{tabular}{c|c|c|c|c|c|c}
    \hline
      \multirow{2}{*}{Module} & \multicolumn{3}{c|}{MSE} & \multicolumn{3}{c}{MAE}   \\ \cline{2-7}
       & 30 min & 60 min & 90 min & 30 min & 60 min & 90 min \\  \hline
       Empty Module        & 1.292 & 3.022 & 4.708 & 0.753 & 1.295 & 1.621 \\  
       Global Branch   & 1.275 & 2.993 & 4.612 & 0.691 & 1.171 & 1.513 \\  
       Detail Branch  & 1.264 & 3.010 & 4.579 & 0.698 & 1.162 & 1.523 \\ 
       Channel Fusion Module   & \textbf{1.197} & \textbf{2.914} & \textbf{4.536} & \textbf{0.681} & \textbf{1.149} & \textbf{1.491} \\ \hline
    \end{tabular}
    }
    \vspace{-0.8cm}
  \end{center}
\end{table}

\section{Conclusion}
\label{sec:typestyle}

The model's ability to extract long-distance correlations significantly influences its prediction accuracy, while dynamically incorporating history and future information aids in comprehending wind speed fluctuation trends. In this context, this paper proposes Windformer, which can effectively extract the long-distance correlation between wind turbines according to the shifted windows and dynamically combine the informative time steps. With this design, Windformer significantly reduces the prediction error and outperforms other current-advanced models.

\vfill\pagebreak

\bibliographystyle{IEEEbib}
\bibliography{strings,refs}

\end{document}